\documentclass[preprint,11pt]{aastex}
\usepackage{amsmath}
\usepackage{threeparttable}
\usepackage{amsmath}
\usepackage{graphicx}
\usepackage{longtable}
\usepackage{natbib}

\DeclareMathSizes{10}{10}{10}{10}

\slugcomment{In Prep. \today}

\shorttitle{Non-Linear Dynamics of Stochastic Disks}
\shortauthors{Cowperthwaite \and Reynolds}

\begin{document}

%% LaTeX will automatically break titles if they run longer than
%% one line. However, you may use \\ to force a line break if
%% you desire.

\title{Non-Linear Dynamics of Accretion Disks with Stochastic Viscosity}

%% Use \author, \affil, and the \and command to format
%% author and affiliation information.
%% Note that \email has replaced the old \authoremail command
%% from AASTeX v4.0. You can use \email to mark an email address
%% anywhere in the paper, not just in the front matter.
%% As in the title, use \\ to force line breaks.

\author{Philip S. Cowperthwaite\altaffilmark{1,2,4} \and Christopher S. Reynolds\altaffilmark{3,4}} 
%\affil{Department of Astronomy, University of Maryland, College Park, MD 20742-2421, USA}
%\affil{Joint Space-Science Institute (JSI), College Park, MD 20742-2421, USA}

%% Notice that each of these authors has alternate affiliations, which
%% are identified by the \altaffilmark after each name.  Specify alternate
%% affiliation information with \altaffiltext, with one command per each
%% affiliation.

\altaffiltext{1}{Harvard-Smithsonian Center for Astrophysics, Cambridge, MD, 02138, USA, pcowpert@cfa.harvard.edu}
\altaffiltext{2}{NSF GRFP Fellow}
\altaffiltext{3}{Department of Astronomy, University of Maryland, College Park, MD 20742, USA}
\altaffiltext{4}{Joint Space-Science Institute (JSI), College Park, MD 20742, USA}

%% Mark off your abstract in the ``abstract'' environment. In the manuscript
%% style, abstract will output a Received/Accepted line after the
%% title and affiliation information. No date will appear since the author
%% does not have this information. The dates will be filled in by the
%% editorial office after submission.

\begin{abstract}
We present a non-linear numerical model for a geometrically thin accretion disk with the addition of stochastic non-linear fluctuations in the viscous parameter. These numerical realizations attempt to study the stochastic effects on the disk angular momentum transport. We show that this simple model is capable of reproducing several observed phenomenologies of accretion driven systems. The most notable of these is the observed linear rms-flux relationship in the disk luminosity. This feature is not formally captured by the linearized disk equations used in previous work. A Fourier analysis of the dissipation and mass accretion rates across disk radii show coherence for frequencies below the local viscous frequency. This is consistent with the coherence behavior observed in astrophysical sources such as Cygnus X-1.

\end{abstract}

%% Keywords should appear after the \end{abstract} command. The uncommented
%% example has been keyed in ApJ style. See the instructions to authors
%% for the journal to which you are submitting your paper to determine
%% what keyword punctuation is appropriate.


\section{Introduction}
Strong and rapid variability is a key observational feature of systems driven by the accretion of material onto a central compact object. The power spectral density (PSD) of emission from such objects is characterized by broadband noise across several decades in frequency. This broadband component is generally well described by a broken-powerlaw form, however the physics that governs such breaks is poorly understood. Additionally, the PSD can feature distinct and resolved peaks indicative of quasi-periodic oscillations (QPOs) in the emission. The broadband noise structure displays many interesting characteristics beyond the PSD. Light curves show log-normality in their structure and a strong linear relationship between the root-mean square (rms) variability and the flux of the source (Uttley \& McHardy 2001; Negoro \& Mineshige 2002; Gaskell 2004). Studies of frequency dependent coherence between energy bands in the time-series data indicates that the emission is coherent at low frequencies then becomes incoherent at high frequencies (e.g. Cygnus X-1, see Nowak et al. 1999 for details). Consequently, many sources also show frequency-dependent phase/time lag behavior between emissions at different wavebands.  This was first clearly demonstrated in accreting stellar mass black holes which often displayed a constant time lag at low frequencies and constant phase at high frequencies (Nowak et al. 1999).  Qualitatively similar behavior is now seen in active galactic nuclei (AGN, Vaughan, Fabian \& Nandra 2003) and accreting non-magnetic white dwarfs (Scaringi et al. 2013).

The broadband variability is often attributed to inward propagating fluctuations driven by stochasticity in the angular momentum transport mechanism (Lyubarski 1997, hereafter L97). Within the framework of the standard Shakura \& Sunyaev (1973) disk model in which the internal ``viscous'' shear stresses are proportional to the pressure ($T_{r\phi}=\alpha P$), L97 attempted to explain this broadband flicker-noise by introducing a small stochastic perturbation into the $\alpha$-parameter. This perturbation generated broadband fluctuations in the accretion ($\dot{m}$).   However, the analysis of L97 was ultimately of a linearized form of the disk equation. In fact, many of the above features, most notably the rms-flux relationship, are not formally captured by this model since linearization means that all perturbations are applied to the base state and do not build upon each other multiplicatively.  Still, it was suggested by Uttley \& McHardy (2001) that the log-normality and rms-flux relation are a result of propagating fluctuations, with the intuitive but formally unproven statement that such features would arise from an analysis of the disk equations with non-linear fluctuations (see also Uttley et al. 2005).

This problem has also been examined by Kelly et al. (2011, hereafter K11) using the tools of stochastic differential equations. Specifically, they studied the use of the stochastic Ornstein-Uhlenbeck (OU) process to model the observed light curves and power spectra of systems with accretion-driven variability. This was motivated by the fact that a solution of the stochastically perturbed linearized disk equation is a linear combination of OU processes which, in the case where the stochastic driving terms are Brownian motion, are simply damped random walks (also see Wood et al. 2001, Titarchuk et al. 2007). K11 found that a linear combination of only one or two such processes was sufficient to well describe the data, and that they could reproduce breaks in the observed power-spectra.  

In this paper, we construct numerical realizations of {\it non-linear} fluctuations in a simple one-dimensional viscous disk.  We show that these disk models produce mass accretion rates and luminosities that feature a linear rms-flux relationship and approximate log-normality.  We also examine the Fourier space relations between the mass accretion rates and dissipations at different radii, one possible proxy for light curves in different energy/wavelength bands.  We find coherence between radii for frequencies less than the viscous frequency.  However, we fail to cleanly reproduce the frequency dependence of the observed time-lags.   Therefore, this toy non-linear model provides some theoretical underpinning for attempts to understand the rms-flux relationship but further developments of the model are required to capture the full behavior of real systems. This work is complimentary to the analysis of the propagating fluctuation model performed by Ingram \& van der Klis (2013), however they begin from a statistical model for the mass accretion whereas we base our analysis on the non-linear disk equations. 

This paper is structured as follows: Section 2 outlines our modifications to the standard viscous disk model and numerical setup, Section 3 presents the results of our simulations, and Section 4 discusses the results of the previous section and provides suggestions for future work. 

\section{Theoretical Basis}

We consider accretion disks in the framework of a simple 1D viscous disk (e.g., see Pringle 1981). We operate within the thin disk approximation which assumes that the material in the disk is narrowly confined in a common orbital plane. This ensures that the vertical scale of the disk (H) is much less than the radial scale (R, $H/R \ll 1$). Consequently, we must also assume that the disk is radiatively efficient. If this were not the case, viscous dissipation would cause the disk to heat to temperatures of order the virial temperature and hence puff up into a quasi-spherical configuration, violating our thin disk assumptions. Furthermore, also required to be consistent with the thin-disk assumption, we assume a fluid element's orbit is essentially Keplerian in nature, with a very small radial velocity responsible for the actual accretion.   While an obvious application of this model is to black hole accretion disks, we neglect relativistic effects and assume a Newtonian gravitational potential ($\Phi=GM/R$).

\subsection{Disk Model}
The canonical equation for such a disk (in units where $G = M = c = 1$) is the non-linear diffusion equation given in Pringle (1981) as
\begin{align}
\frac{\partial \Sigma}{\partial t} = \frac{3}{R} \frac{\partial}{\partial R}\left [ R^{1/2} \frac{\partial}{\partial R}(\nu \Sigma R^{1/2}) \right ],\label{eq:basiceqn}
\end{align}
where $R$ is the distance from the central engine, $\Sigma(R,t)$ is the surface density of the disk material, and $\nu(R,t)$ is the effective shear viscosity of the disk.  As we discuss in detail below, $\nu$ will be taken as a stochastic variable with some (prescribed) stationary base state $\nu_0(R)$ to which we add spatially-uncorrelated but temporally-correlated stochastic fluctuations.    We now wish to make solutions to eqn.~(\ref{eq:basiceqn}) more computationally convenient with this eventual application in mind.    First, we introduce the change of variables $x = R^{1/2}$. Equation~(\ref{eq:basiceqn}) then becomes
\begin{align}
\frac{\partial \Sigma}{\partial t} = \frac{3}{4x^3} \frac{\partial^2}{\partial x^2}(\nu \Sigma x).
\label{eq:mixeq}
\end{align}

Multiplying both sides of Equation~\ref{eq:mixeq} by the term $\nu_0 x$ (which has no $t$ dependence) while defining $\Psi_0 \equiv \nu_0 \Sigma x$ and $\Psi \equiv \nu \Sigma x$, equation (2) can be written as
\begin{align}
\frac{\partial \Psi_0}{\partial t} = \frac{3 \nu_0}{4x^2} \frac{\partial^2 \Psi}{\partial x^2},
\label{eq:}
\end{align}
which is the form of the disk evolution equation that serves as the basis for our numerical solution.

We are primarily concerned with the local mass accretion rate across, and viscous dissipation at, each radius.  The instantaneous mass accretion rate is
\begin{align}
\dot{m}(R,t) = 3\pi R^{1/2}\frac{\partial}{\partial R}(R^{1/2} \nu \Sigma),
\end{align}
which in our new variables translates to
\begin{align}
\dot{m}(x,t) = 3\pi \frac{\partial \Psi}{\partial x}.
\end{align}

We can perform a similar manipulation of the dissipation,
\begin{align}
D(R,t) = \frac{1}{2} \nu \Sigma \left (R \frac{d\Omega}{dR} \right )^2 = \frac{9 \nu \Sigma}{4 R^3},
\end{align}
which gives
\begin{align}
D(x,t) =  \frac{9  \Psi}{4x^7}.
\end{align}
The total disk luminosity is then given by
\begin{align}
L_{\text{disk}}(t) = \int_{\text{disk}} [D(R,t)] (2 \pi R) dR = \int_{\text{disk}} \frac{9 \Psi}{x^4} dx.
\end{align}

\subsection{Stochastic Model}
We now assume a stochastic perturbation in the viscous parameter of the form $\nu = \nu_0 (1 + \beta)$.  Here, $\beta(x,t)$ is the fluctuation relative to the baseline level $\nu_0(x)$, and physically represents the turbulent fluctuations in the Maxwell stresses of a magnetohydrodynamic accretion disk. Unlike previous treatments (L97 and K11), we do not assume that $\beta$ is small.  To proceed, we need to specify the nature of the random process $\beta$.    We employ assumptions similar to those of L97.   Specifically, we suppose that the fluctuations are spatially uncorrelated (likely to be valid for $\Delta R>H$) but have temporal correlations on some timescale $\tau$.  Thus, we impose the condition that $\langle \beta(R_1,t)\beta(R_2,t)\rangle \rightarrow 0$ when $R_1 \ne R_2$. The characteristic timescale imposes the condition that $\langle \beta(R,t_1)\beta(R,t_2)\rangle \rightarrow 0$ when $\left \vert t_1 - t_2 \right \vert /\tau \gg 1$. 

A specific realization of these conditions can be generated by an Ornstein-Uhlenbeck (OU) process. The OU process, models the evolution of a statistical variable of interest as it responds to an additive input noise with an exponential decay to its mean (see Kelly et al. 2009, K11). This evolution is governed by a stochastic differential equation of the form
\begin{align}
d\beta(t) = - \omega_0(\beta(t) - \mu)dt + \xi dW(t), \quad \omega_0 > 0,
\label{eq:ou}
\end{align}
where $\omega_0$ is the characteristic frequency, $\mu$ is the mean of the process, $\xi$ is the driving noise amplitude, and $W$ denotes a Wiener process.  In the typical case where $W$ is Brownian motion, $dW$ is Gaussian white noise.  Equation.~\ref{eq:ou} then simply represents a damped random walk.

Equation~\ref{eq:ou} can be made physically relevant for our disk with the appropriate choice of parameters. We assume the mean in the fluctuations is zero (since the average component of the viscosity is already described via the $\nu_0$ term.).   Next, we take guidance from L97 and, for most of the work presented here, assume that the relevant timescale for the fluctuations is the local viscous timescale such that $\omega_0(x) = \omega_a(x) = 1 / \tau_a(x)$, where $\tau_a(x) = x^2/\nu_0$. We take $dW$ to be Gaussian white noise with zero mean and unity standard deviation. 

To ensure that the rms amplitude of the fluctuations $\beta$ are independent of radius we note that when $dW$ has zero mean the variance is given by K11 as $\xi^2 \tau/2$. Therefore, we must take $\xi$ as proportional to the square root of the local accretion time. We normalize this factor to the accretion time at the inner edge of the disk yielding $\xi = \sqrt{\omega_a(x) / \omega_a(x_0)}$.

The final form of the OU equation is then:
\begin{align}
d\beta(x,t) = - \omega_a(x)\beta(x,t)dt + \sqrt{\frac{\omega_a(x)}{ \omega_a(x_0)}}dW(x,t),
\label{eq:ou2}
\end{align}
which satisfies all of the physically relevant requirements for the distribution of fluctuations.   

We need one additional restriction on the viscous fluctuations in order to make our mathematical model well-posed. If, due to a downward fluctuation, the overall viscosity $\nu$ becomes negative over an extended region of the disk, the basic diffusion equation is unstable and singularities will develop in the solution.  Thus, we impose a positivity criterion --- whereever and whenever the solution of eqn.~\ref{eq:ou2} results in $\beta<-1$ (and hence $\nu<0$), we peg $\beta=-1$. Lastly, when evolving $\Psi$, we scale the amplitude of $\beta$ by a fiducial value of 0.5. This is done to control the effect of high sigma spikes on the stability of the solution.   

\subsection{Numerical Considerations}
We performed our simulations using an explicit numerical integration algorithm. The simulation was run on a spatial grid with $x_{\rm in} = 1$ and $x_{\rm out} = 100$ with $\Delta x = 0.1$. We chose a fiducial value for the baseline viscosity $\nu_0=10^{-3}$. Simulations were run with a time-step computed to safely satisfy the necessary CFL stability conditions, typically $\Delta t = 0.2$. The initial condition on $\psi_0$ was chosen such that the disk began in a steady state with unity accretion rate, i.e. $\partial^2 \Psi_0 / \partial x^2 = 0 \; \Rightarrow \; \Psi_0(x,t=0) = (x - x_{\rm in})/3\pi$.  Lastly, we imposed Dirchlet boundary conditions $\Psi(x_{in},t) = \Psi_0(x_{\rm in},t=0)$ and $\Psi(x_{\rm out},t) = \Psi_0(x_{\rm out},t=0)$. In order to protect the simulation from stochastic effects at the outer boundary we established a buffer zone between $x=95$ and $x=100$ where $\beta = 0$. The simulation was run for a duration of $t_{\rm max}=3\times 10^7$, and the state of the disk was recorded with a cadence of $\Delta t_{out} = 100$.   

\section{Results}
\begin{figure*}[!h]
   \centering
\epsscale{1}
\plotone{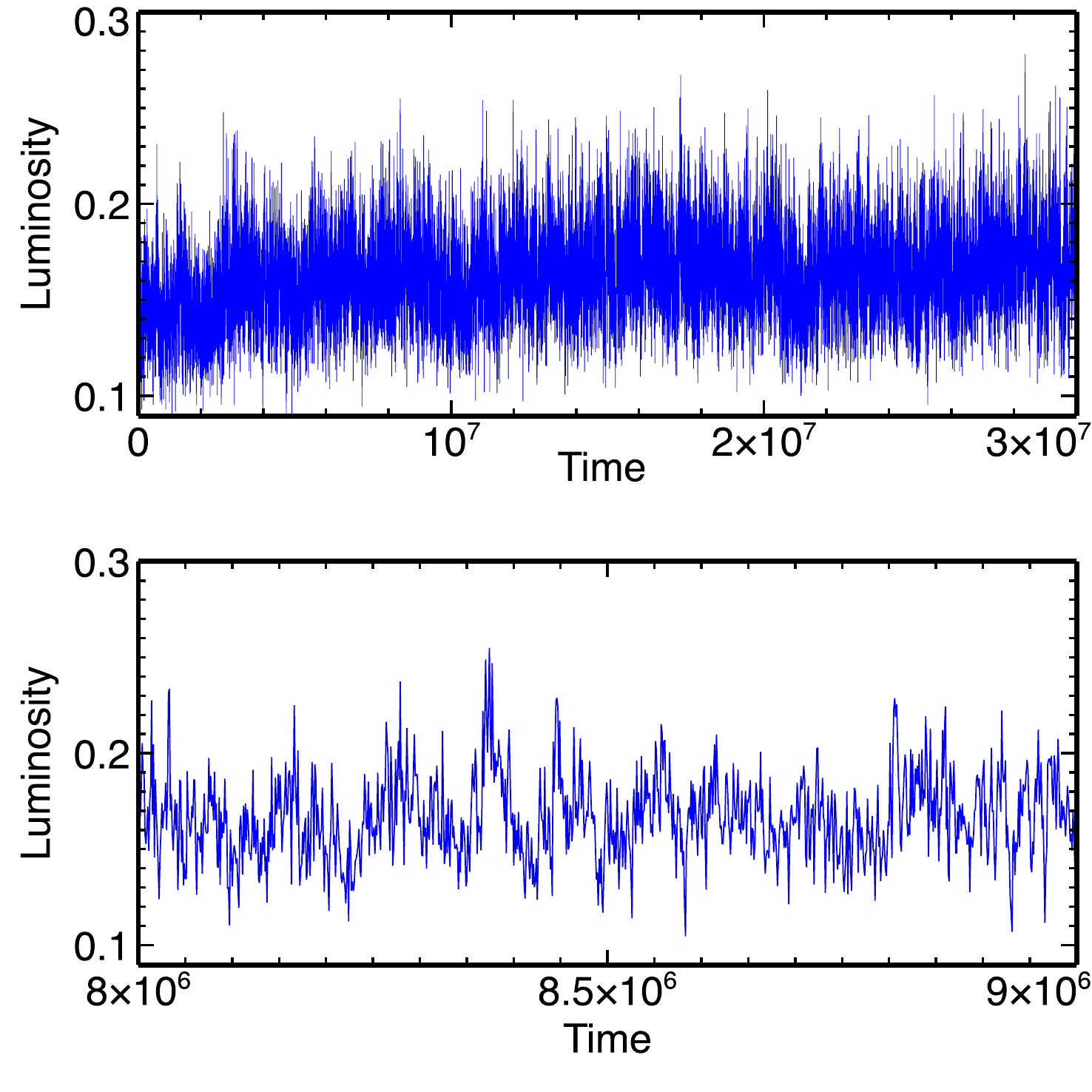}
   \caption{The synthetic light curve as derived from the total disk luminosity. The top panel shows the entire duration of the simulation. The bottom panel shows an arbitrary 10$^6$ M region (i.e the length of a Fourier segment, see Section 3.2 for details). For clarity, every 100th data point is plotted. The qualitative behavior in both plots, specifically the decrease in fluctuation amplitude during dips and flares relative to the baseline, hints at log-normal behavior.}
   \label{fig:lightcurve}
   \end{figure*}

Our primary goal is to examine whether the phenomenology of the broadband noise in real sources can be reproduced by our toy model. Consequently, we focus our analysis on the dissipation of the disk as this most closely corresponds to the observational emissions of real astrophysical disks. We integrate the dissipation across the entire disk at each time to produce the total time-dependent luminosity of the disk.  This light curve can be seen in Figure~\ref{fig:lightcurve}. Qualitatively, it resembles the light curve for an accreting source. Log-normality can be visually inferred by observing the decreased amplitude of fluctuations during dips compared with the flares. 

\begin{figure*}[!h]
   \centering
\epsscale{1}
\plotone{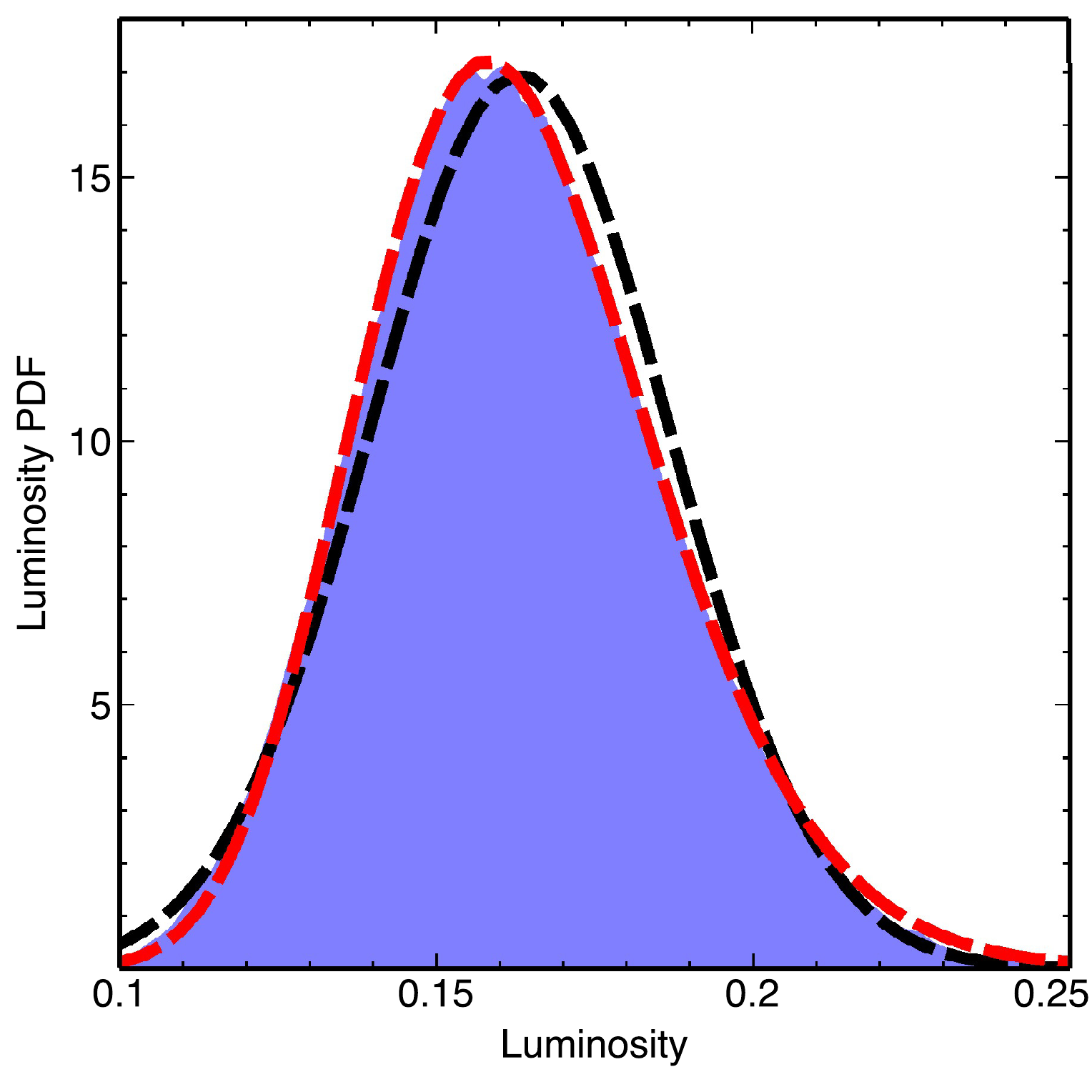}
   \caption{The shaded blue region shows the synthetic probability distribution function for the total disk luminosity. The dashed red line indicates the best-fitting log-normal distribution while the black dashed line indicates the best-fit normal distribution. While neither is statistically preferable, the overall disk behavior suggests the log-normal distribution is more appropriate.}
   \label{fig:distribution}
   \end{figure*}

The luminosity distribution is shown in Figure~\ref{fig:distribution}. We fit this distribution with a standard log-normal function of the form 
\begin{align}
f(L ; \mu , \sigma) = \frac{1}{\sqrt{2 \pi} L \sigma} \exp \left [-\frac{(\ln(L) - \mu)^2}{2 \sigma ^2} \right ],
\end{align}
where $L$ is the luminosity and the fitting parameters $\mu$ and $\sigma$ are the mean and standard deviation respectively. We found best fit parameters of $\mu = -1.84\pm0.04$ and $\sigma = 0.22\pm0.02$. It must be noted that we also found a reasonable, albeit visually slightly worse, fit using a normal distribution with $\mu = 0.16\pm0.02$ and $\sigma = 0.03\pm0.01$. The best-fit lines are plotted in Figure~\ref{fig:distribution}. 

%The distribution is noisy due to the fact that the simulation has only run 30 times longer than the viscous timescale out the outer radius, leaving significance ``cosmic variance'' in the lightcurve.  This variance makes a $\chi^2$ comparison between the distribution and the log-normal/normal less meaningful. However a K-S test between the observed distribution and the log-normal fit returned P = 0.008, while a similar test using the normal fit returned P=0.003. Thus, there is weak statistical evidence that the log-normal distribution is preferred. It is likely that longer runs will be necessary to comment on the distribution with any statistic certainty. However, the existence of an rms-flux relationship established in the next section serves as strong evidence for the log-normality of the underlying luminosity distribution.

\subsection{The Rms-Flux Relationship}
\begin{figure*}[!h]
   \centering
\epsscale{1}
\plotone{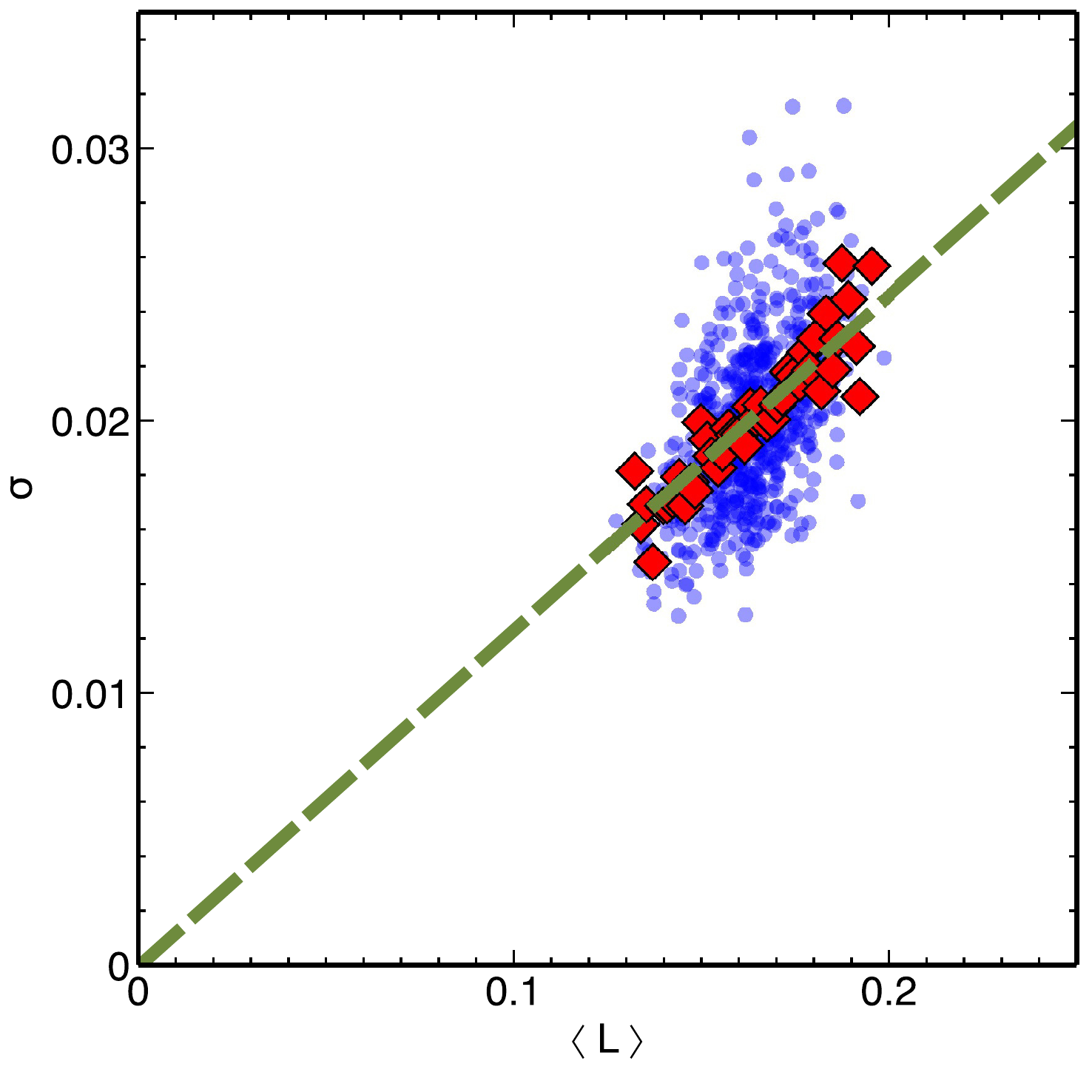}
   \caption{The root-mean square variability as a function of the average luminosity.  The blue circles show the unbinned data while the red diamonds show the binned data. The green dashed line indicates the best linear fit to the binned data. The linear relationship between the rms variability and average luminosity is a strong phenomenological feature of observed sources and not traditionally captured by the disk equations.}
   \label{fig:rmsflux}
   \end{figure*}

We investigated the existence of an rms-flux relationship in our data using the integrated luminosity as a function of time. The resulting light curve was divided into 600 equal segments of $N=50,000$. For a black hole of 10$M_{\odot}$ this corresponds to $\sim2.5$\,s per segment. The mean luminosity $\langle L \rangle$ was computed for each segment. We then computed the rms fluctuation in each segment as 
\begin{align}
\sigma^2 =  \frac{1}{N} \sum_{i = 1}^N (L_i - \langle L \rangle)^2,
\end{align}
where $L_i$ is the luminosity at the $i^{\text{th}}$ point in the segment. The resulting rms-flux plot can be seen in Figure~\ref{fig:rmsflux}. 

Due to the nature of the stochastic noise in the simulation it is necessary to bin the resulting rms-flux data. The data were linearly binned into 50 segments and then averaged over each bin. The linear nature of the binned data is immediately apparent. Following the prescription of Uttley \& McHardy (2001) we fit the binned data with a line of the form $$\hat{\sigma} = k(\langle L \rangle + C)$$ where $k$ and $C$ are fitting constants. We obtained a fit to the data with $k = 0.12\pm0.02$ and $C = (5\pm1)\times10^{-4}$. The fitted line is shown on Fig.~\ref{fig:rmsflux} to trace the data qualitatively well. The fact that our simulation produces a linear result is an important confirmation of the log-normality of the light curve. Furthermore, this directly supports the phenomenological model that the multiplicative effect of numerous inward propagating fluctuations are the cause of the observed log-normality.

\subsection{Fourier Analysis of Light Curves}
\begin{figure*}[!h]
   \centering
\epsscale{1}
\plotone{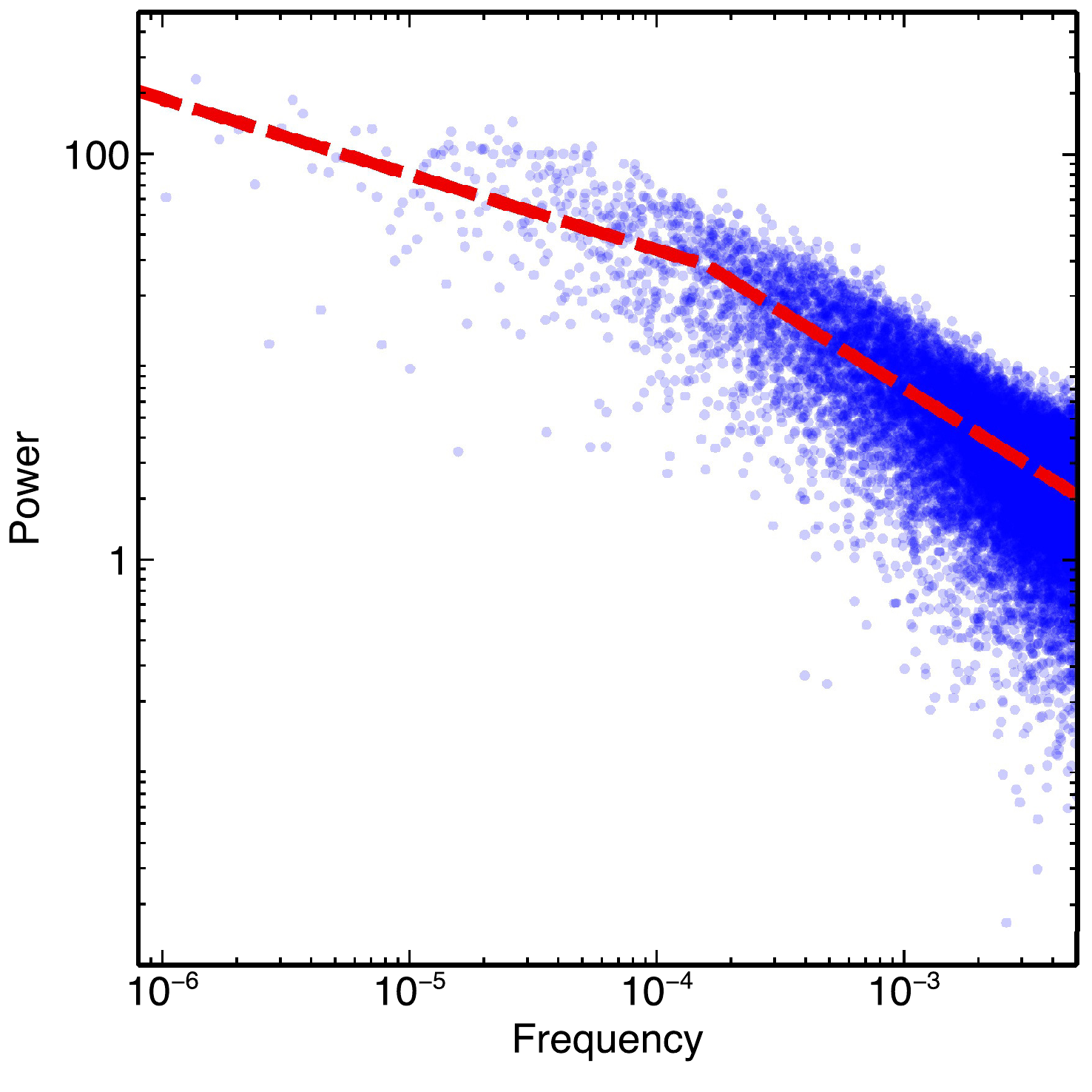}
   \caption{The power spectrum of the total integrated disk luminosity. The red dashed line indicates a broken power-law fit to the data. There is a clear break in the power spectrum around $f \sim 10^{-4}$. This is consistent with the viscous frequencies across the inner disk regions}
   \label{fig:psd}
   \end{figure*}

We next wish to understand how our light curves behave in frequency space. We divided the total integrated disk luminosity into 36 segments, each of length 819200 M ($\sim$ 40 seconds at 10 $M_{\odot}$). The fast Fourier transform was taken for each segment.  We used these Fourier segments, denoted $P_i(f)$, to construct the power spectrum density (PSD) of the $i^{\text{th}}$ segment as per the standard prescription $$ \langle |P_i(f)|^2 \rangle = \langle P_i^{\star}(f) P_i(f) \rangle $$ where the use of angle brackets denotes an averaging over segments. The resulting power spectrum across the inner boundary is shown in Figure~\ref{fig:psd}. 

It can be seen that the PSD is consistent with a system dominated by broadband noise. There are no QPOs, as expected given the simple nature of the model. We attempted to fit the PSD to a broken power-law of the form $P(f) \propto (f/f_{\text{th}})^{-\zeta_1},\; f < f_{\text{th}}$ and $P(f) \propto (f/f_{\text{th}})^{-\zeta_2},\; f > f_{\text{th}}$, where we take the threshold frequency to be a free parameter of the model. The best-fit returned $\zeta_1 = 0.37\pm0.01$ and $\zeta_2 = 0.76\pm0.01$ at a break frequency of $f_{\text{th}} = (1.5\pm0.1)\times10^{-4}$. The slope is likely not consistent with the $\zeta = -1$ expectation put forth by L98 because we are not dealing with a strict $\alpha = \text{const.}$ accretion disk. However, while the power-law dependence is flatter than that indicated by observational data, this indicates a clear break in the power spectrum as expected. The break frequency is comparable to the viscous frequency for the inner regions of the disk (i.e. $x < 5$). Given that the dissipation is dominated by the inner regions it is reasonable to expect the PSD to obey these timescales. 

\subsubsection{The Coherence Function}
A key aspect of the propagating fluctuation model is the behavior of the fluctuations as they propagate inward from the outermost radii. Therefore, it is important to analyze to what extent various disk radii are capable of communicating and at what frequencies this communication may occur. This can be accomplished through the use of the coherence function. From Nowak et al. (1999) we define the coherence function, $\gamma(f)$, as follows: consider two time series $s$ and $h$ with Fourier transforms $S$ and $H$. The coherence function between $s$ and $t$ is then $$\gamma^2(f) = \frac{|\langle S^{\star}(f)H(f)\rangle |^2}{\langle |S(f)|^2 \rangle \langle |H(f)|^2 \rangle}$$ where, again, the angle brackets indicate an averaging over segments of the data. It then becomes clear that the coherence function is merely the modulus of the averaged cross-spectrum between two time series normalized by the averaged power spectra of the individual series. Consequently, the coherence function takes on values between zero and unity. 

\begin{figure*}[!h]
   \centering
\epsscale{1}
\plotone{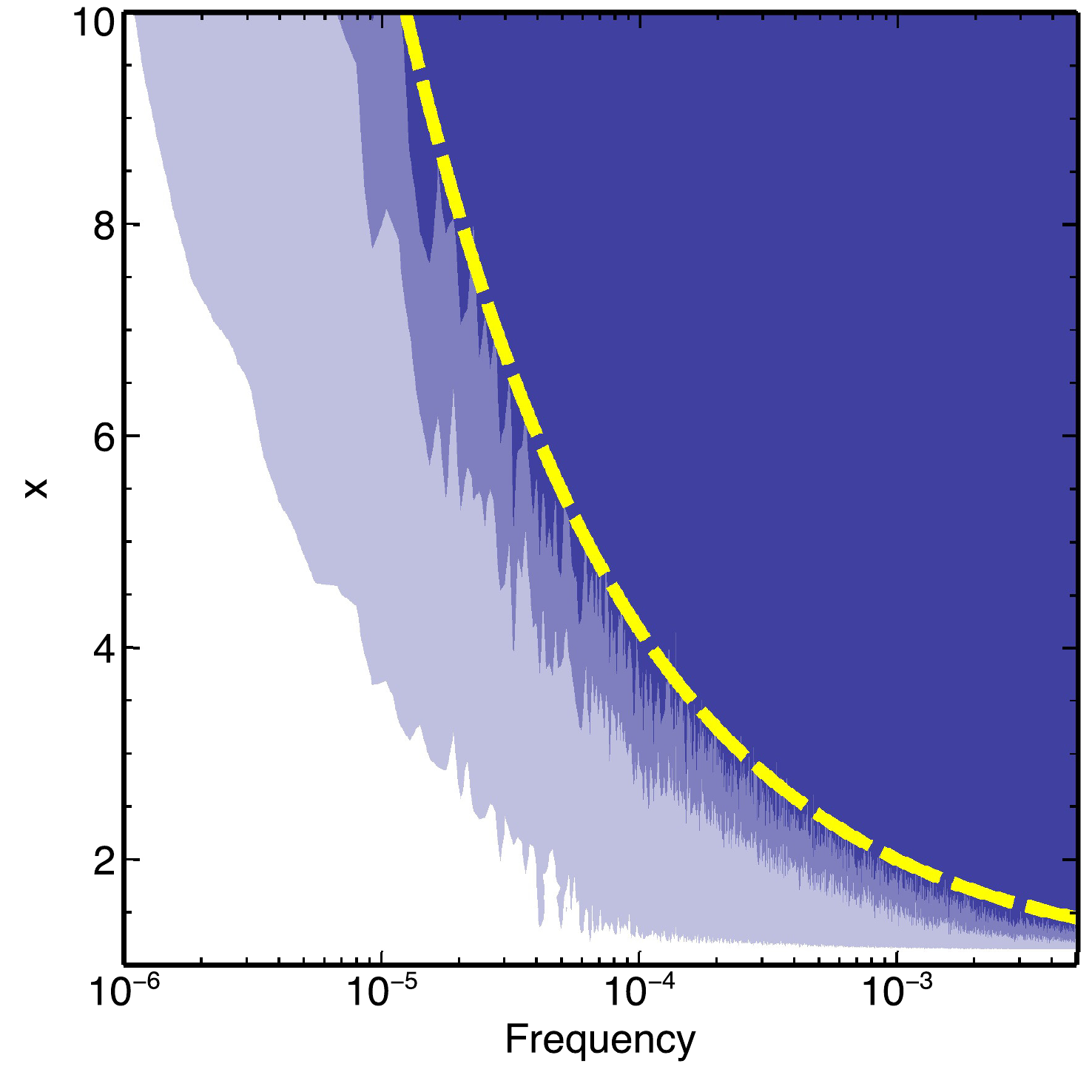}
   \caption{The coherence map for the dissipation across the inner radii of the disk. The shading indicates regions of $\gamma^2 \in [0.6, 0.98]$, $\gamma^2 \in [0.3, 0.6]$, and $\gamma^2 < 0.3$, from lightest to darkest respectively (i.e white indicates $\gamma^2 > 0.98$). The dashed yellow line indicates the local viscous frequency. It can be seen that radii become incoherent as perturbations fall below this frequency.}
   \label{fig:coherencemap_visc}
   \end{figure*}

In the case of our data we take the dissipation at different radii as our time signals. We can choose the inner radius as the first time series and compute its coherence function with all of the other radii in order to produce a coherence map, $\gamma^2(R,f)$. Such an approach can be thought of as being analogous to the common practice of studying the coherence function between different wavebands. If we assume the emission is locally a blackbody then the natural temperature gradient in the disk will cause emission from two radii to occur at different wavebands. The resulting coherence map for the disk as shown in Figure~\ref{fig:coherencemap_visc}. 

There is clear structure in the map as radii become coherent on roughly the viscous frequency or less. This provides some indication of the timescales on which radii communicate information about the perturbations. We should note that coherence does not necessary arise because the radii are globally in perfect phase. Rather, it indicates that a linear transformation exists between the time series in the time domain, i.e., 
\begin{align}
h(t) = \int_{-\infty}^{+\infty} t_r(t - \tau)s(\tau)d\tau,
\end{align} 
where $t_r$ is the transfer function (Nowak et al. 1999). Our coherence map confirms the notion of L97 that, at a given radius, fluctuations are passed down to smaller radii only if they fall below the local viscous frequency.

\subsubsection{Phases and Lags}
If two radii of the disk have time-coherent dissipations, this implies the existence of well-defined (frequency-dependent) phase shifts between the dissipation time-series.   More precisely, if we write the cross-spectrum between two time series as a complex number then we have $\langle S^{\star}(f) H(f) \rangle = A(f) e^{- \phi(f)}$ where $\phi(f) = \phi_S(f) - \phi_H(f)$ is the frequency-dependent Fourier phase shift. The calculation of the phase shift produces values in the domain from $[-\pi , \pi]$ with the assumption that $\phi(f) \rightarrow 0$ as $f \rightarrow 0$. We can now scale the phase by the frequency yielding the frequency-dependent time lag: $\tau(f) = \phi(f) / 2 \pi f$. Given the nature of our model, the Fourier time lags represent the physical reprocessing time of the fluctuations as they are diffused in the disk. 

\begin{figure*}[!h]
   \centering
\epsscale{1}
\plotone{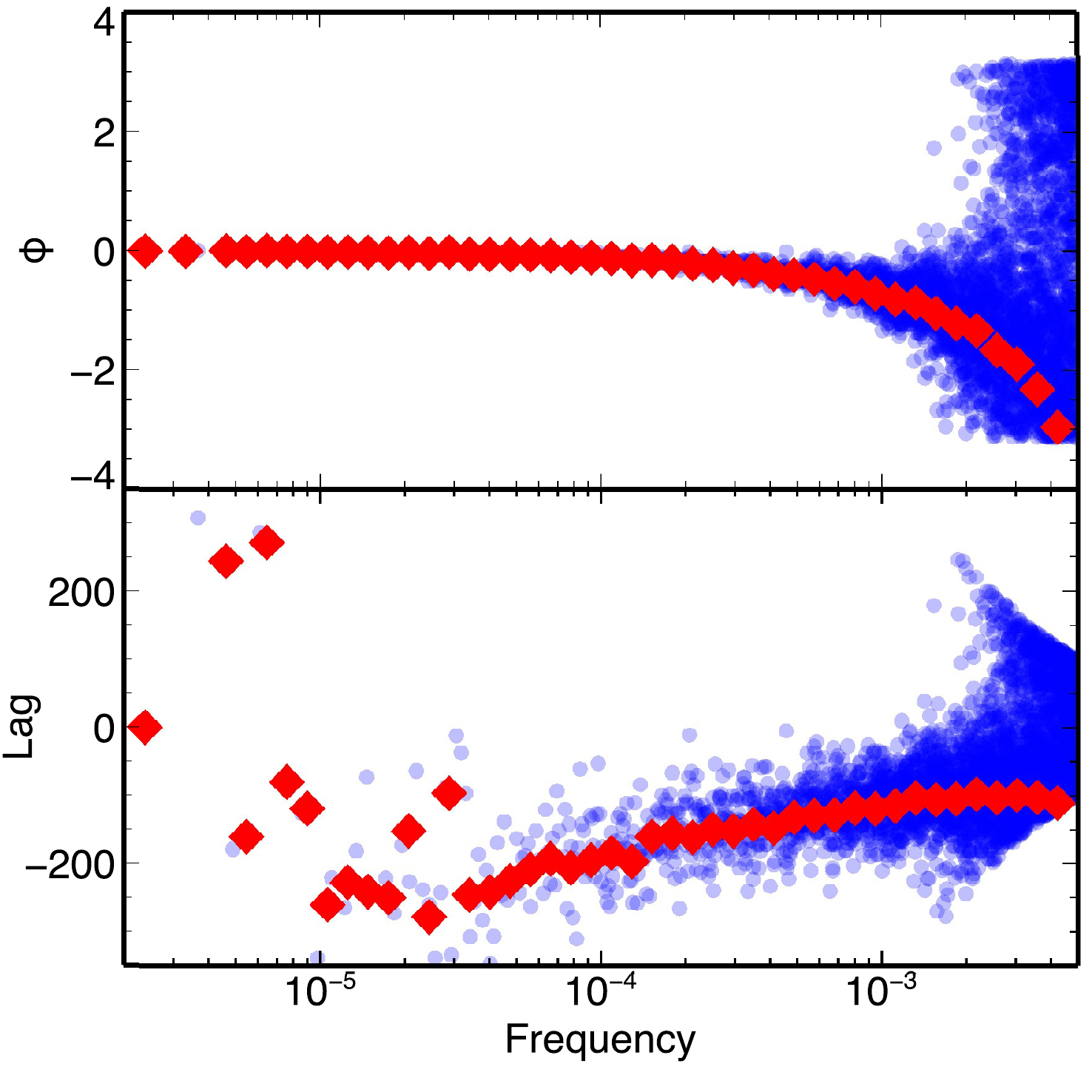}
   \caption{The cloud of blue circles show the unbinned data while the red diamonds show the binned data. The top panel shows the Fourier phase of the cross-spectrum between the dissipation at $x=1$ and $x=2$. The bottom panel shows the associated time lag. }
   \label{fig:phase_1_2}
   \end{figure*}

To illustrate the behavior of the phases and time lags, Fig.~\ref{fig:phase_1_2} shows these quantities derived from the dissipation at $x=1$ and $x=2$ (corresponding to $R=1$ and $R=4$).  From the coherence map (Fig.~\ref{fig:coherencemap_visc}), we can see that these two radii are extremely coherent at low frequencies ($\gamma^2>0.98$ for $f<3\times 10^{-4}$), and then the coherence gradually falls to low values ($\gamma^2<0.3$) over a span of about 1.5 decades of frequency around the viscous frequency.  Consistent with this, at low frequency there is little scatter in the phases from the individual segments.  The mean phase gradually increases magnitude as frequency increases in a manner consistent with approximately equal time-lag.   We can see that at frequencies around the local viscous frequency ($f_{visc} \sim 10^{-3}$) the scatter in the phase increases (corresponding to the decreasing coherence).   At frequencies above $f\approx 2\times 10^{-3}$, the scatter in phase becomes large and phases get wrapping, becoming almost randomly scattered between $-\pi$ and $\pi$ at $f=(3-4)\times10^{-3}$.  The sharp decrease in the averaged phase shift above this frequency is entirely an artifact of averaging incoherence processes with a phase wrapped cross spectrum argument. In order to counteract some effects of phase wrapping we binned the data at the level of the cross-spectrum. The phases and lags were then computed from these binned data.

Looking at the time lag information across the same radii we see that the binned lag data shows a lag which is small for very low frequencies, increases in magnitude to $|\tau| \approx 150$ at $f\approx 10^{-4}$ and then very slowly declines in magnitude to $|\tau| \approx 200$ before coherence is lost.  Given our convention, the sense of this lag is that the inner ($x=1$) dissipation profile lags the dissipation further out, expected from inwardly propagating fluctuations.  At frequencies above the local viscous frequency the lag begins to weakly rise before the radii become incoherent.   By comparison, studies of the Fourier lag in Cygnus X-1 have shown lags with a strong frequency dependent decline ($f^{-1}$ and hence consistent with approximately constant phase shift; Nowak 1999). This is clearly not visible in the dissipation data. It is interesting to note that the lag times are much less than the naively computed local viscous timescale, $t_{visc} \sim 1000$.   

\begin{figure*}[!h]
   \centering
\epsscale{1}
\plotone{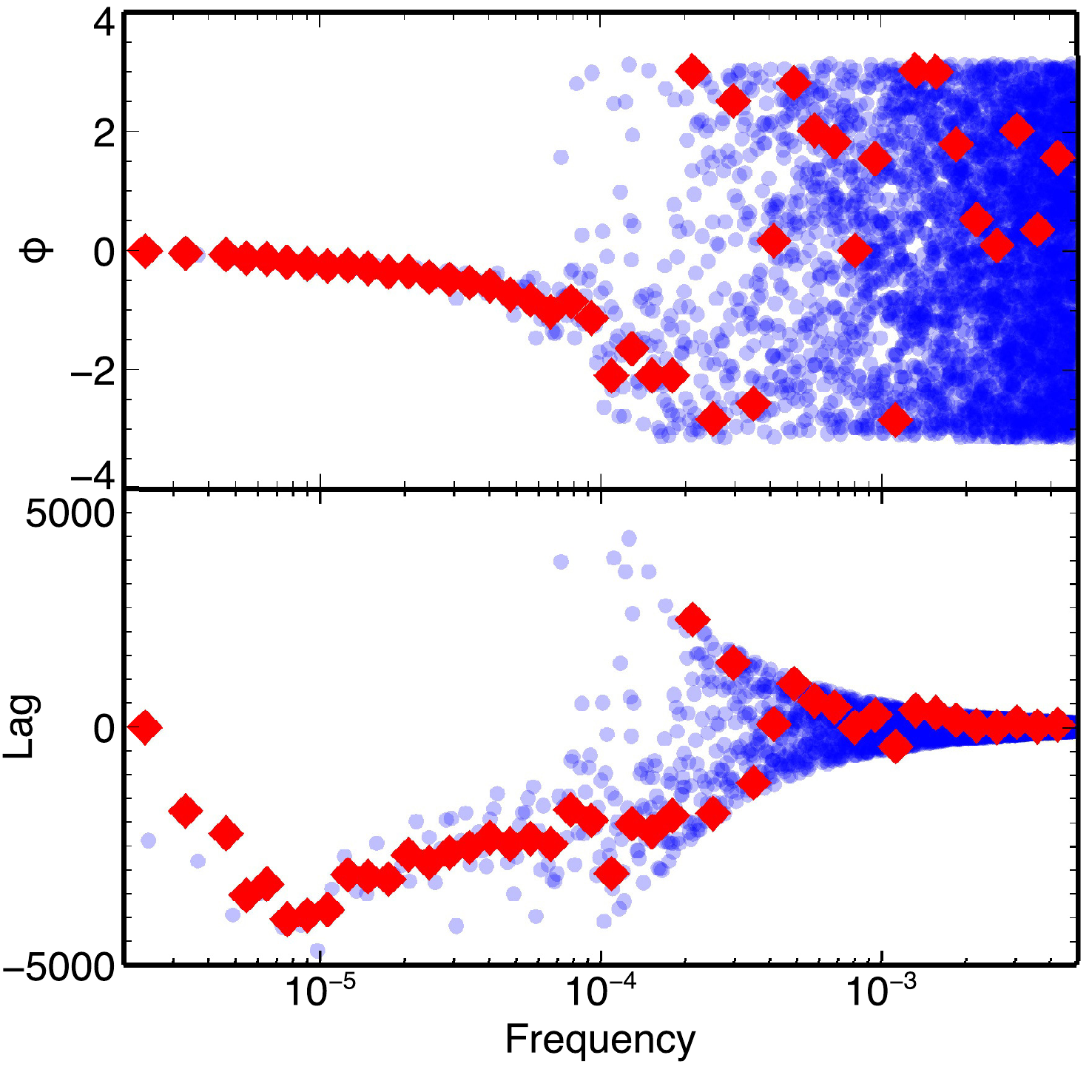}
   \caption{As Figure~\ref{fig:phase_1_2}. The top panel shows the Fourier phase of the cross-spectrum between the dissipation at $x=2$ and $x=5$. The bottom panel shows the associated time lag.}
   \label{fig:phase_2_5}
   \end{figure*}

When looking at the innermost radii it is important to check for the possible effects of the computational boundary on our analysis. We repeated the Fourier analysis, this time computing the coherence between the dissipation at radii $x=2$ and $x=5$.   The viscous frequency between these two radii is $f = 1.1\times10^{-4}$.   We see in Fig.~\ref{fig:phase_2_5} an interesting pattern of time lags, with slowly decreasing lags below the viscous frequency, and negligible lags (approximately constant phase) above the viscous frequency.  Coherence is completely lost (resulting in essentially randomly distributed phases) for $f>10^{-3}$.  This pattern of time lags has tantalizing similarities to the behavior seen in Cygnus~X-1 and other accreting sources.

\subsubsection{Intradisk Behavior}
The above analysis considers the behavior of the dissipation across the entire disk. We also investigated the local behavior of key observables (e.g. light curves, histograms, PSDs, and the rms-flux relationship) at the specific radii $x = 2$, $x=5$ and $x=10$. We focused on the inner radii as that is the region of the disk where the dissipation is most dominant. Qualitatively, we observe the same behavior at each radii as the disk shows globally. We observe dissipation light curves and histograms with indications of log-normal behavior, PSDs that show a clear break, and a linear rms-flux relationship.

\subsection{Differences Between $\dot{m}$ and L}
While we have considered the observational features of our simulated disk it is important to consider other physical processes. Given its dynamical importance, there is motivation to study the behavior of $\dot{m}$ in our simulations. Repeating the above analysis for $\dot{m}$ we found many of the same behavioral properties. There is clear log-normality in the light curve along with the expected linear rms-flux relationship. The PSD is composed entirely of broadband noise and shows no break at the viscous frequency at the inner radius.

\begin{figure*}[!h]
   \centering
\epsscale{1}
\plotone{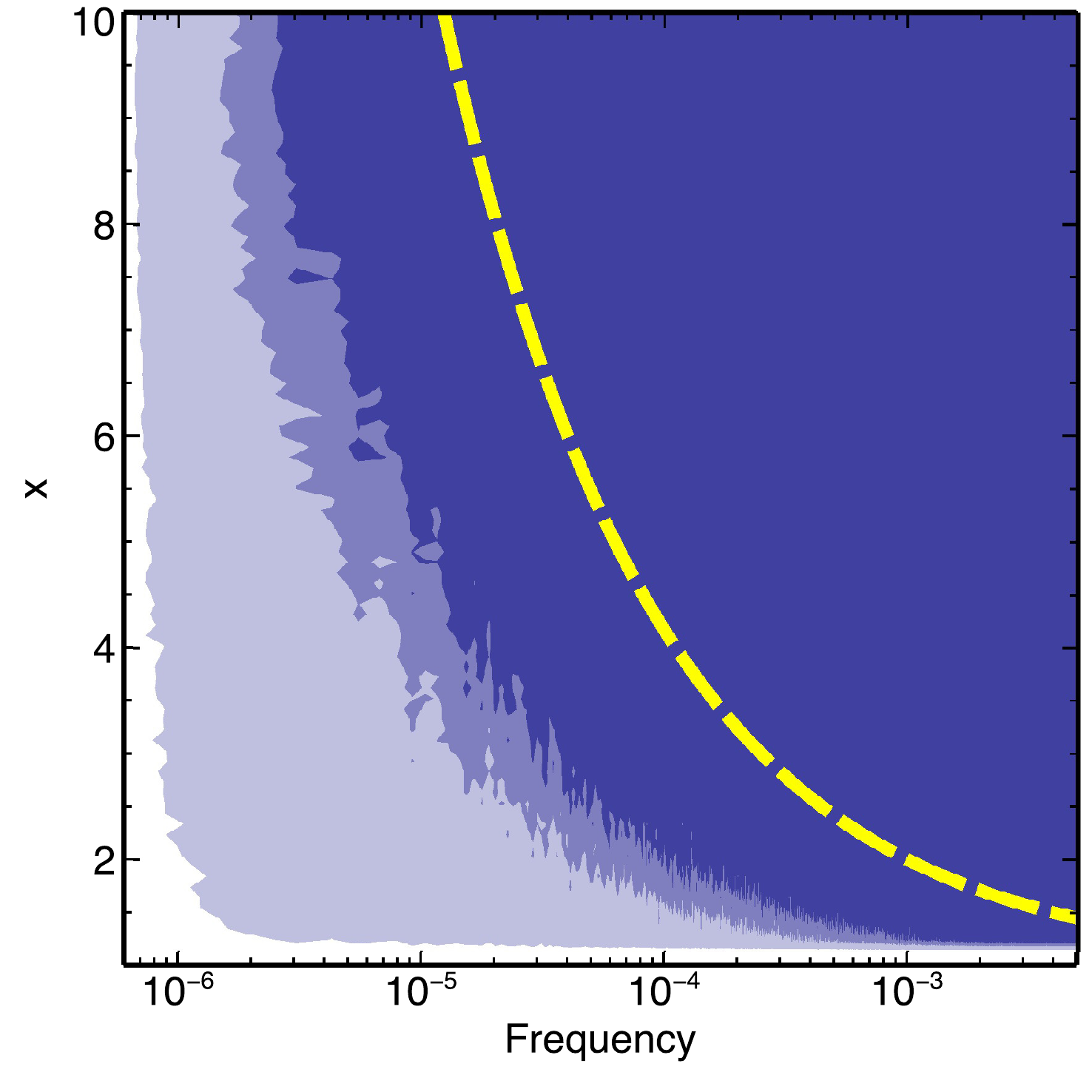}
   \caption{The coherence map for $\dot{m}$ across the inner radii of the disk. The shading indicates regions of $\gamma^2 \in [0.6, 0.98]$, $\gamma^2 \in [0.3, 0.6]$, and $\gamma^2 < 0.3$, from lightest to darkest respectively (i.e white indicates $\gamma^2 > 0.98$). The dashed yellow line indicates the local viscous frequency. We can see that $\dot{m}$ becomes coherent at lower frequencies compared to the dissipation. }
   \label{fig:coherencemap_mdot}
   \end{figure*}

\begin{figure*}[!h]
   \centering
\epsscale{1}
\plotone{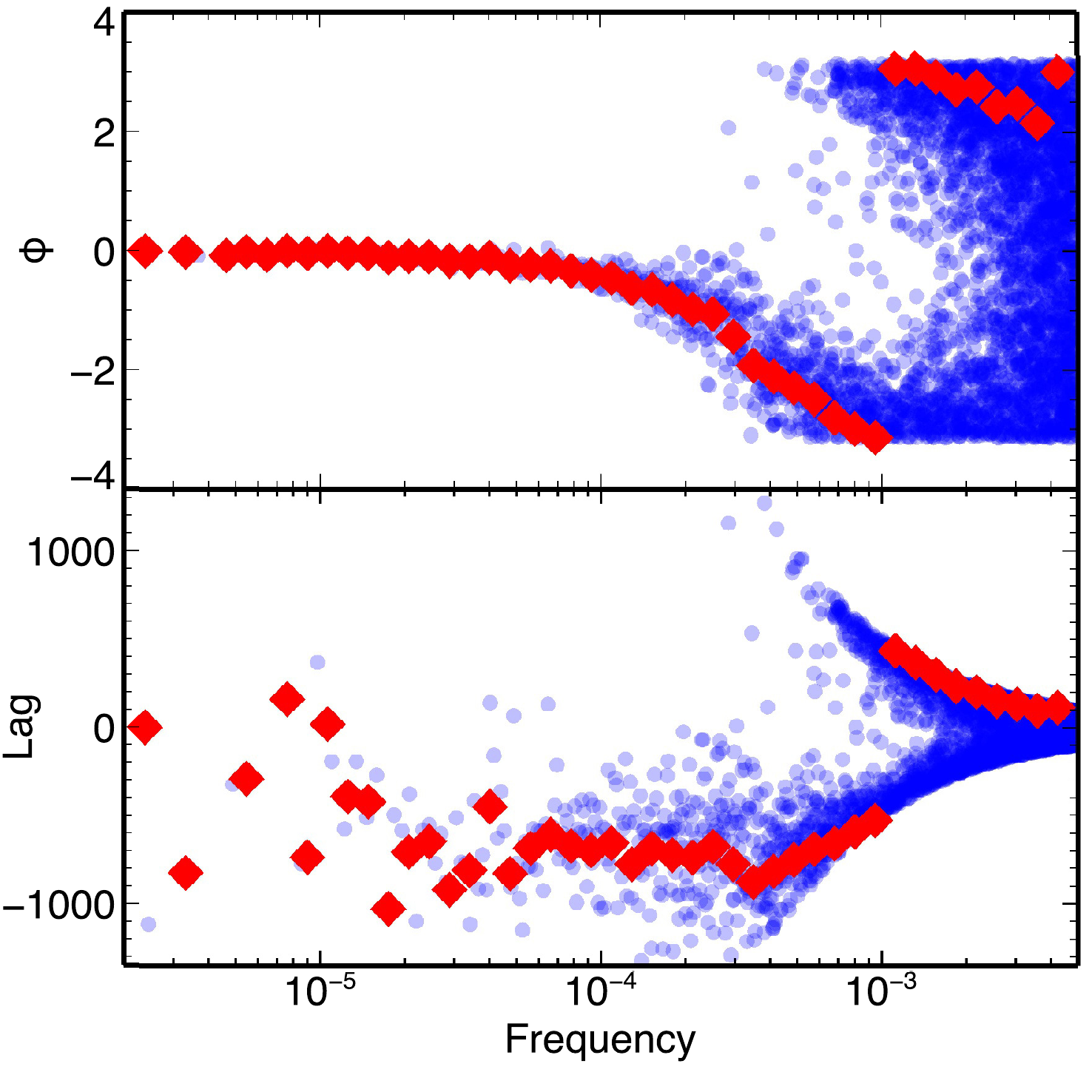}
   \caption{As Figure~\ref{fig:phase_1_2}. The top panel shows the Fourier phase of the cross-spectrum between $\dot{m}$ at x=1 and x=2. The bottom panel shows the associated time lag. The behavior around $10^{-3} < f < 2\times10^{-3}$ may be attributable to ``phase-wrapping." }
   \label{fig:phase_mdot}
   \end{figure*}

The most striking differences between the two datasets are the coherence map and subsequent phase/lag information. The coherence map shows the same functional form as the as the dissipation, however $\dot{m}$ appears to become coherent only at lower frequencies when compared to the dissipation (Fig.~\ref{fig:coherencemap_mdot}). We can understand this by examining the phases/lags at two illustrative radii; again we choose $x=1$ and $x=2$.  We see in Fig.~\ref{fig:phase_mdot}, for $f<4\times10^{-4}$, there is a constant time lag.  However, this lag is substantially larger than that found for the dissipations, $|\tau|\approx 600-800$ as opposed to the $|\tau|\sim 150$ for the dissipations.  It is clear from Fig.~\ref{fig:phase_mdot} that the corresponding larger phase shift starts to phase wrap well before there is a loss of coherence.  Thus, the coherence function $\gamma^2(f)$ is suppressed at intermediate frequencies by phase wrapping well before there is a genuine loss of communication between the radii.  

%Looking at the Fourier lag plot, we see that the system is in a state of constant lag with $t \sim 500$ M before showing the expected convergence to zero as the radii become incoherent. We note the behavior in the ``phase-wrapped" region does begin to resemble some of the phenomenological behavior seen in the time lags of real objects.

\begin{figure*}[!h]
   \centering
\epsscale{1}
\plotone{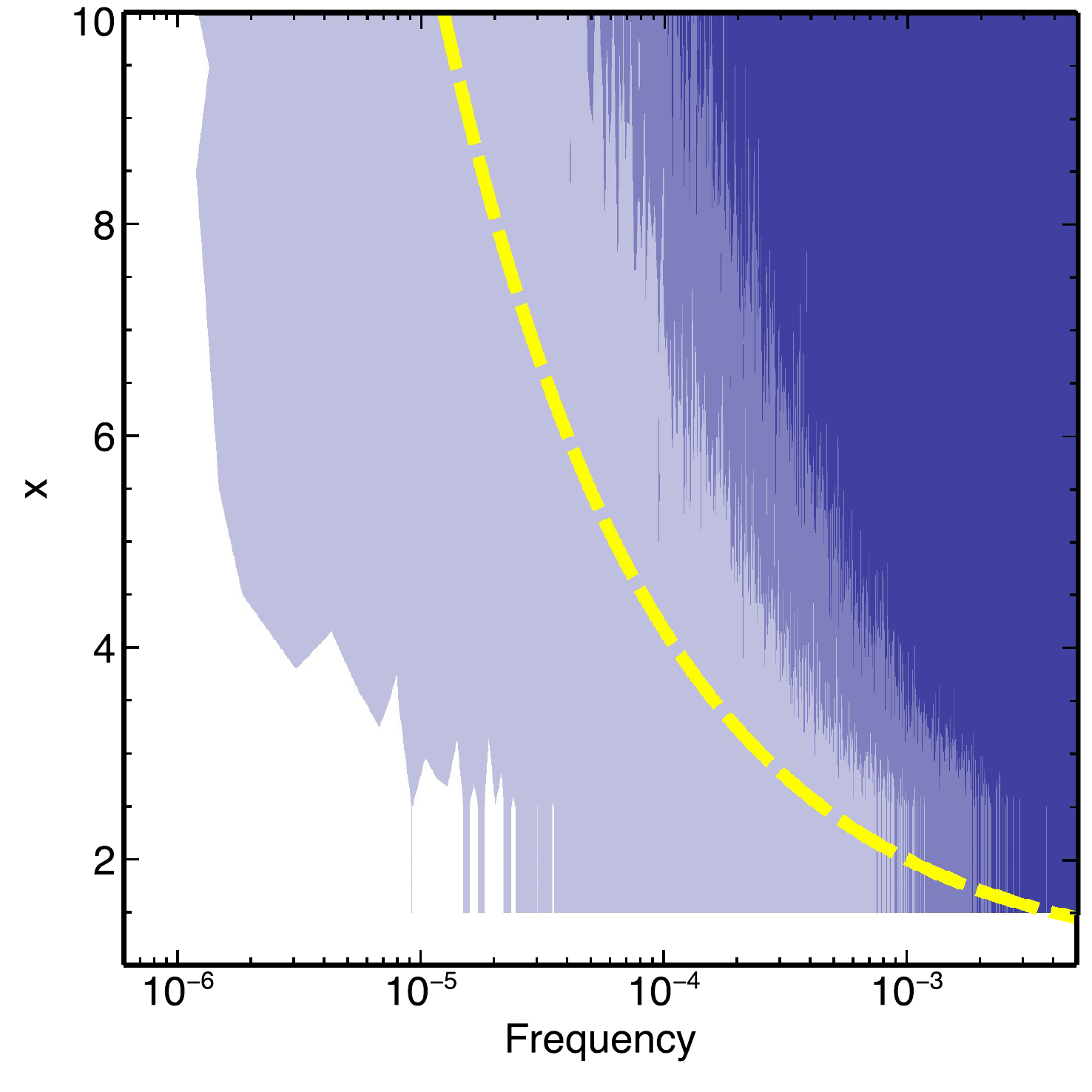}
   \caption{The coherence map for $\dot{m}$ vs. dissipation across the inner radii of the disk. The shading indicates regions of $\gamma^2 \in [0.6, 0.98]$, $\gamma^2 \in [0.3, 0.6]$, and $\gamma^2 < 0.3$, from lightest to darkest respectively (i.e white indicates $\gamma^2 > 0.98$). The dashed yellow line indicates the local viscous frequency. We can see that the observed behavior is closer to that of $\dot{m}$ than the dissipation.}
   \label{fig:coherencemap_mdot_diss}
   \end{figure*}

\begin{figure*}[!h]
   \centering
\epsscale{1}
\plotone{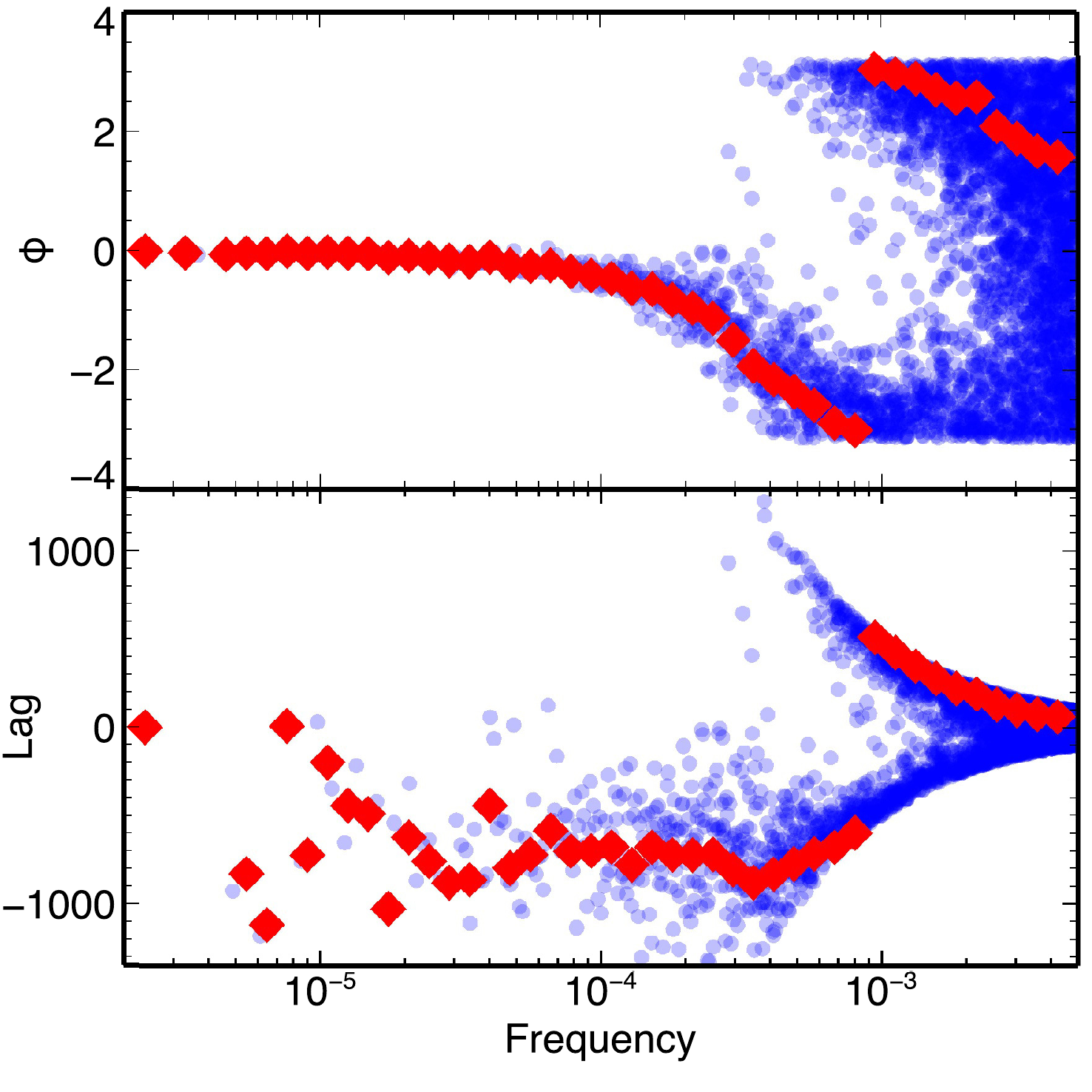}
   \caption{As Figure~\ref{fig:phase_1_2}. The top panel shows the Fourier phase of the cross-spectrum between $\dot{m}$ and dissipation at x=1 and x=2. The bottom panel shows the associated time lag. The behavior at $f > 10^{-3}$ is clearly attributable to ``phase-wrapping." }
   \label{fig:phase_mdot_diss}
   \end{figure*}

For completeness, we examine the Fourier relationship between the dissipation and mass accretion rate.  This could be astrophysically relevant if, for example, the hard X-ray emission from an accreting black hole was tied closely to mass accretion rate through the innermost circular stable orbit but the soft X-rays were thermal emission from a dissipative disk.   Looking at the coherence map, we see that the regions of high coherence again trace the viscous frequency with regions becoming coherent at frequencies similar to those seen for $\dot{m}$ (Figure~\ref{fig:coherencemap_mdot_diss}).  Interestingly, there is evidence of convergence to a constant phase of $\sim \pi$ (which is also phase wrapped to $-\pi$) in this phase plot (Figure~\ref{fig:phase_mdot_diss}).

\subsection{Considering Alternate Timescales}
While the viscous timescale is one natural choice as the characteristic frequency for the driving perturbations it may be illuminating to consider other possibilities. The only other relevant quantity in our toy model is the dynamical frequency, $f_{\text{dyn}}=x^{-3}$, assuming Keplerian orbits. We ran a second simulation with the characteristic frequency of the OU process set to the dynamical frequency. Analysis of the resulting data showed a disk that was strongly inconsistent with the aforementioned phenomenologies. While the simulation was able to produce a light curve with log-normality and a linear rms-flux relationship, the disk was not able to produce reasonable behavior in frequency space. 

\begin{figure*}[!h]
   \centering
\epsscale{1}
\plotone{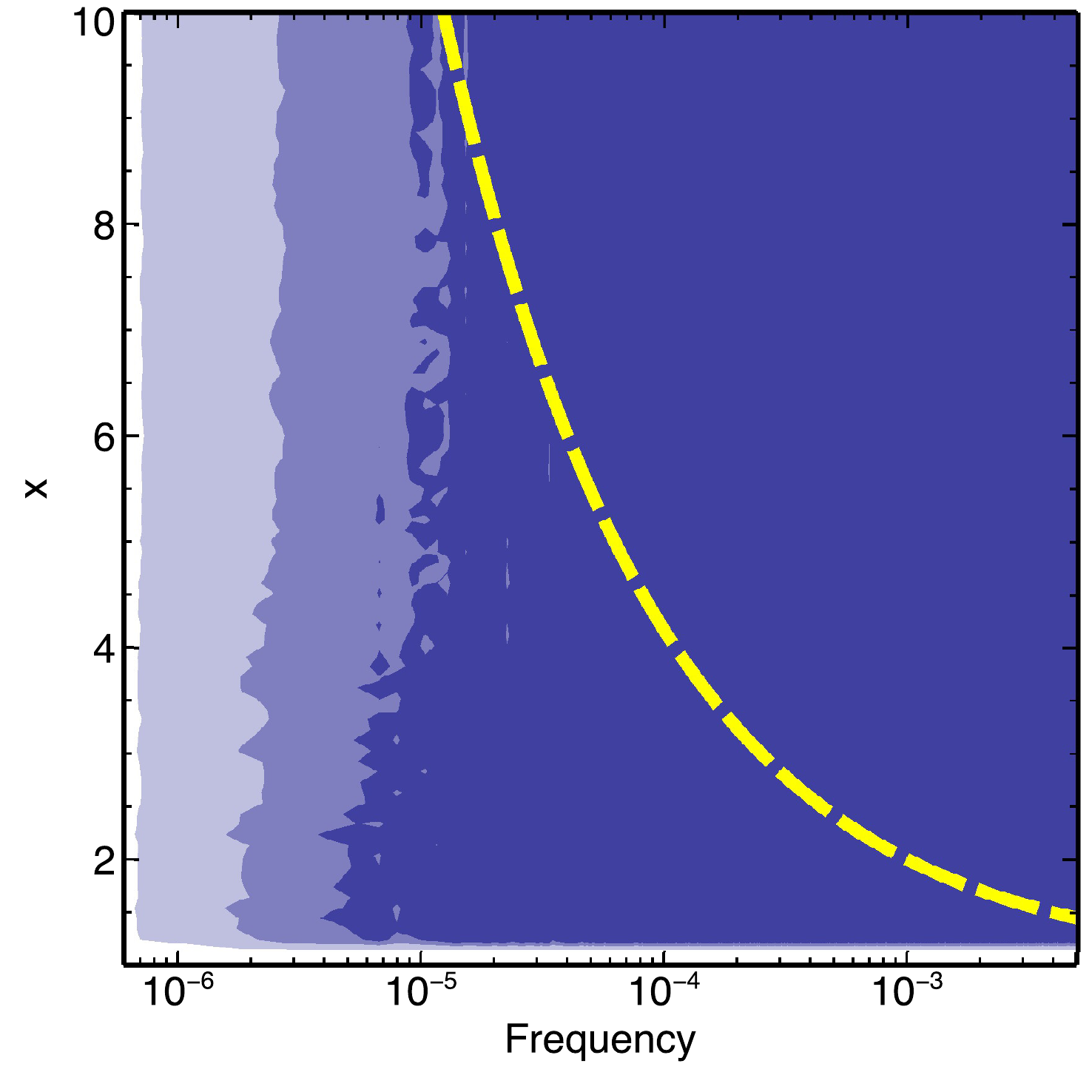}
   \caption{The coherence map for dissipation across the inner radii of the disk with perturbations driven by the local dynamical frequency. The shading indicates regions of $\gamma^2 \in [0.6, 0.98]$, $\gamma^2 \in [0.3, 0.6]$, and $\gamma^2 < 0.3$, from lightest to darkest respectively (i.e white indicates $\gamma^2 > 0.98$). The dashed yellow line indicates the local viscous frequency. We can see that there is coherence at the lowest frequencies but otherwise there is no discernible structure.}
   \label{fig:coherencemap_dyn}
   \end{figure*}

\begin{figure*}[!h]
   \centering
\epsscale{1}
\plotone{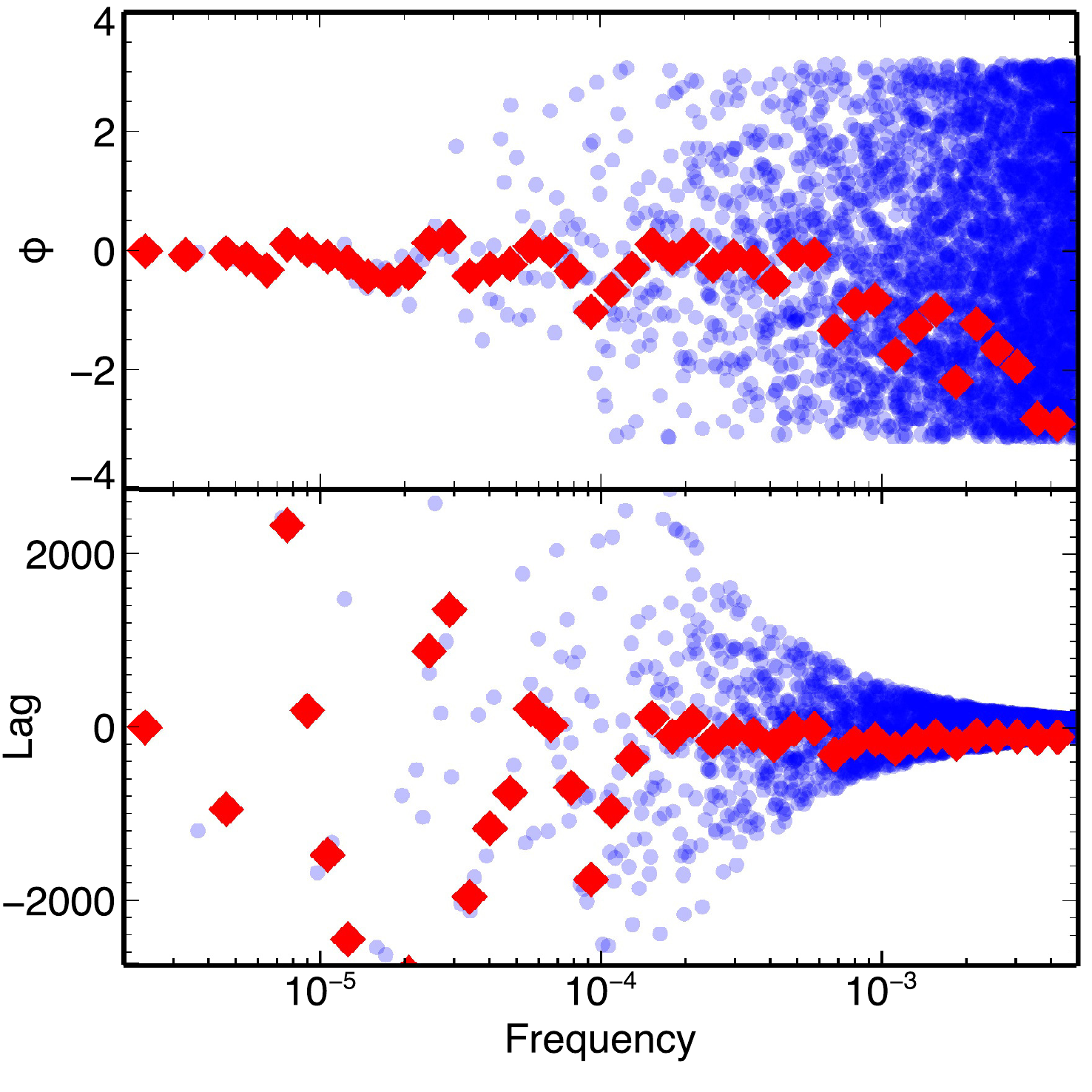}
   \caption{As Figure~\ref{fig:phase_1_2}. The top panel shows the Fourier phase of the cross-spectrum between the dissipation at x=1 and x=2. Here the perturbations are governed by the dynamical timescale. The bottom panel shows the associated time lag. The observed envelope structure in the lag plot is merely $\pi f^{-1}$ due to the random scattering of phases.}
   \label{fig:phase_dyn}
   \end{figure*}

The coherence map for both the dissipation and the mass accretion rate show large regions of little to no coherence, while the regions of high coherence are concentrated at low frequencies or along the inner boundary. As an example, the coherence map for the dissipation is show in Figure~\ref{fig:coherencemap_dyn}. There is no hint of any meaningful structure in these plots. Due to lack of general coherence we do not expect to see any frequency-dependent phase behavior apart from at the lowest possible frequencies. The phase and lag plots are shown in Figure~\ref{fig:phase_dyn}. We do note that there is a weak trend in the binned data above the viscous frequency likely due to a combination of noise and diffusion effects in the disk.

\section{Discussion and Conclusion}
We have presented one of the first numerical attempts to understand the ``propagating fluctuation" model of Lyubarskii (1997) in the context of non-linear accretion disk models. While the observational effects of this model have long been intuited from detailed studies of observational data, many key phenomenological features are not rigorously attributable to the linear model. One of the most striking features is the linear relationship between the rms flux-variability and the average flux. We have shown that our toy model, which starts with the basic disk equations, is capable of reproducing this relationship.  

Going beyond the rms-flux relationship, we can examine the behavior of our model disk in Fourier space.  We find that the time-series of the dissipations at two different radii have temporal coherence up to frequencies comparable to, or slightly greater than, the viscous frequency relating the two radii.  Across most of this range of coherence frequencies, the relationship between the dissipations appears to be one of constant time lag, with the inner radius lagging the outer radius by about a tenth of a viscous timescale.  At higher frequencies there is some hint, seen most clearly when comparing radii away from the inner computational boundary, for a constant phase regime (with a time-lag that correspondingly declines with increasing frequency), but this only lasts over about a factor of $\sim 3$ in frequency before all coherence is lost.   

This frequency behavior in real sources has been well studied observationally.   For example, detailed studies of Cygnus~X-1 with the {\it Rossi X-ray Timing Explorer (RXTE}; Nowak et al. 1999; Poutanen 2000; Gilfanov et al. 2000; Uttley et al. 2011) reveal coherence between the soft and hard x-ray bands at frequencies below the viscous frequency, possible regions of constant time-lag, and extended regions of approximately constant phase shift.   Our model is able to reproduce the coherence behavior, but does not show extensive regions of constant phase. Nowak et al. 1999 also noted that black hole X-ray binaries show ``stepped" lag behavior, with the appearance of several sub-components each with their own constant lag. It was thought that these steps in lag were due to the Lorentzian sub-components used to model the PSD. It may then be the case that our observed constant lag behavior is the result of a single Lorentzian component seen when comparing specific pairs of radii (i.e. $x=1$ and $x=2$), whereas in observations of an astrophysical disk we are observing the contributions of many radii.

Of course, given the simplicity of our model and treatment, the inability to explain the detailed phenomenology of real sources is not surprising.  Observationally, coherence and time-lag studies must be performed on lightcurves extracted in particular bands.   Even in the case where the observed emission is thermal, a given waveband picks out emission from a broad range of radii and the detailed lightcurve may be poorly approximated by the dissipation at one radius. In addition, of course, it is possible for one (or even both!) of the energy bands under study to be well above the thermal energy of the disk and, hence, actually originating from an X-ray corona --- the uncertain relationship between the underlying state of the disk ($\dot{m}$ or dissipation) and the emission from the X-ray corona is another barrier in matching this simple model to the observations.   Future developments of our approach will include more realistic models for relating the disk dynamics to the observables.

Ultimately, real astrophysical accretion disks are extremely complex systems. Even our 1D model may miss key disk behaviors by assuming azimuthal symmetry. For example, the work of Dexter \& Agol (2011) showed that 2D accretion disks with stochastic temperature fluctuations end up strongly inhomogeneous. However, it is not clear if their stochastic behavior is driven by stochasticity in the accretion flow as captured by our model or is yet another complication in understanding the behavior of the accretion disk. Clearly, the full range of interactions present require 3D magnetohydrodynamical models, often with general relativistic effects and radiative transfer, to fully understand.   However such models are computationally expensive, and for the foreseeable future it will not be feasible to run a GRMHD model for many outer disk viscous timescales.   Thus, simple models such as that presented in this paper can be viewed as a bridge between the full GRMHD simulations and the observations, into which lessons from the simulations (such as the statistics of the fluctuations or the physics of the emission processes) can be imported. Future development of these models will follow such a direction.

We thank Phil Uttley for helpful discussion about this work. We also thank our anonymous referee for their helpful comments about this work. PSC is grateful for support provided by the NSF through the Graduate Research Fellowship Program, grant DGE1144152. CSR thanks NASA for funding under the Astrophysical Theory Program grant NNX10AE41G.

\end{document}